\documentclass{PoS}

\usepackage{slashed}
\usepackage{units}

\newcommand{\beq}{\begin{equation}}
\newcommand{\eeq}{\end{equation}}
\newcommand{\bit}{\begin{itemize}}
\newcommand{\eit}{\end{itemize}}
\newcommand{\bea}{\begin{eqnarray}}
\newcommand{\eea}{\end{eqnarray}}
\newcommand{\st}{\ensuremath{\tilde{t}_1}}
\newcommand{\neutr}{\ensuremath{\tilde{\chi}_1^0}}
\newcommand{\charg}{\ensuremath{\tilde{\chi}_1^\pm}}
\newcommand{\met}{\ensuremath{\slashed{E}_T}}
\newcommand{\mpt}{\ensuremath{\slashed{p}_T}}

\title{Light stop phenomenology}

\ShortTitle{Light stop phenomenology}

\author{\speaker{Jong Soo Kim}\\
       ARC Centre of Excellence in Particle Physics at the Terascale, \\
       School of Chemistry and Physics, \\
       The University of Adelaide, Adelaide SA 5005 Australia\\
       E-mail: \email{jongsoo.kim@adelaide.edu.au}}

\author{Manuel Drees\\
        Bethe Center for Theoretical Physics, Bonn\\
        E-mail: \email{drees@th.physik.uni-bonn.de}}

\author{Marja Hanussek\\
        Bethe Center for Theoretical Physics, Bonn \& SISSA/ISAS, Trieste\\
        E-mail: \email{hanussek@th.physik.uni-bonn.de}}

\abstract{We consider the discovery potential of light stops in the MSSM at the LHC. Here, we assume that the lightest neutralino is the LSP and that the lighter stop is  the NLSP. Direct stop pair production is difficult to probe in scenarios with a small mass splitting between the stop and a neutralino. We discuss two different search channels: the monojet and the two $b$--flavoured jets and large missing transverse energy signature. We present the discovery reach in the stop--neutralino mass plane for both channels. The latter process is sensitive to the stop--higgsino--$b$ quark coupling. This allows us to test a supersymmetry relation involving superpotential couplings. We briefly comment on the possible precision with which the coupling can be measured.}

\FullConference{36th International Conference on High Energy Physics,\\
		July 4-11, 2012\\
		Melbourne, Australia}

\begin{document}

\section{Introduction}
Recent results from the ATLAS search \cite{ATLAS_search} suggest that first generation squarks and gluinos below 1.5 TeV are excluded if their masses are equal. This might already be considered somewhat unnatural, since TeV scale supersymmetry has been introduced to stabilise the electroweak hierarchy. However, at one loop order, the leading radiative corrections to the Higgs mass depend only on third generation sparticles. In this talk, we consider the direct production of stops. The production cross section is much smaller compared to the ones for first generation squarks and gluinos and thus the analyses published by ATLAS and CMS are not so competitive compared to the results from squark and gluino searches. As long as the mass difference between the stop and the lightest supersymmetric particle (LSP) is large enough, the search strategies are still based on multi--jet and missing transverse momentum signatures. In addition, $b$--quarks or lepton final states are included in these searches \cite{Aad:2012yr}. However, if the mass splitting becomes small, the decay products of the stop becomes very soft and thus the amount of missing transverse momentum is considerably reduced. We assume that the right--handed stop is the only light strongly interacting particle with a mass difference between the stop and the bino--like neutralino LSP of a few (tens of) GeV. The chargino and all the other neutralinos are heavier than the stop and thus the stop dominantly decays into a charm and a neutralino LSP. We discuss two possible search channels for such a scenario. One of the search channels renders the possibility to estimate the size of the coupling between a higgsino--like chargino, a stop and a $b$ quark, which allows us to check a supersymmetry (SUSY) coupling relation involving Yukawa couplings for the first time. We conclude with a brief discussion of the prospects of testing this particular SUSY coupling relation.

\section{Monojet Signature}
First, we want to consider stop pair production in association with one hard QCD jet \cite{Carena:2008mj,Drees:2012dd}.
\beq \label{proc}
pp\rightarrow \st\st^* j + X,
\eeq
where $X$ is the rest of the event. We study scenarios with mass differences between the lighter (mostly right--handed) stop and lightest bino--like neutralino LSP not larger than a few tens of GeV. In such a framework, on--shell decays such as $\tilde t_1  \rightarrow \tilde \chi_1^+ b$ and $\st \rightarrow b \neutr W$ are kinematically closed and four body decays like $\tilde t_1 \rightarrow \tilde \chi_1^0 \ell^+ \nu_\ell b$ are strongly phase space suppressed. However, the stop decay into the lightest neutralino and a charm jet,
\beq \label{stdec}
\st\rightarrow c\neutr\,,
\eeq
is kinematically open assuming that the $\tilde c$ component of the $\st$ is non--vanishing \cite{Hikasa:1987db,Muhlleitner:2011ww}. Due to the small mass splitting between the stop and the neutralino, both charm jets in the signal will be rather soft on average. The charm jets are then not useful for suppressing the SM backgrounds, since soft jets with fake missing transverse energy are ever-present at the LHC. Thus our collider signature is a single high $p_T$ jet balanced by large missing transverse energy,
\beq \label{signal}
pp\rightarrow j\met.
\eeq
accompanied by additional jets from initial state radiation and the $\st$ decay products. 
With increasing stop masses, the QCD bremsstrahlung becomes stronger, while with larger mass difference between $\st$ and $\tilde\chi_1^0$, the charms can be reconstructed and thus we can also have multiple jets with large missing transverse energy in the final state. However, in a large region of parameter space, our topology is quite simple compared to standard supersymmetric collider signatures: a single energetic jet, which is back to back to the missing transverse momentum vector.

We neglect supersymmetric backgrounds since we assume that all other coloured sparticles are quite heavy which seems to be consistent with recent null results from SUSY searches at the LHC. If the charm jets in the signal are very soft, the dominant irreducible SM background is given by $Z(\rightarrow \nu \bar\nu)+j$. Fortunately, its size can be measured from $Z(\rightarrow \ell^+ \ell^-)+j$ data.  The remaining backgrounds are given by $W(\rightarrow \ell \nu) + j$, $W(\rightarrow \tau \nu) + j$ and $t\bar t$, which are much smaller in size. 
We neglect the single top background, since it only contributes to $1\%$ of the total SM background. 
We do not consider pure QCD dijet and trijet production, since it is expected that a large \met cut removes those backgrounds \cite{atlasConf}. 
We also neglect gauge boson pair production as background, since the total cross section is much smaller than that for single gauge boson plus jet production.

We impose the following cuts: we require one hard jet with $p_T\ge500$ GeV. We further suppress the SM backgrounds by demanding $\met>450$ GeV. We veto all events with a reconstructed electron or muon with $|\eta|<2.5$ We only include isolated electrons with $p_T>$ 10 GeV, but all muons with $p_T>4$ GeV. Events with an identified tau jet with $|\eta|<2.5$ and $p_T>20$ GeV are also vetoed. We require a veto on all tagged $b$--jets with $p_T>20$ GeV and $|\eta|<2.5$ and veto the existence of a second jet with $p_T>100$ GeV. 
More details about the numerical analysis and the numerical tools used in this study can be found in Ref. \cite{Drees:2012dd}. 

In Fig. \ref{fig:mstop_mneut_distribution} we show the statistical significance $S/ \delta B$ in the stop--neutralino mass plane for an integrated luminosity of 100 fb$^{-1}$ at $\sqrt{s}=14$ TeV. The significance of the signal $S$ depends on the error $\delta B$ of the background which is given by
\bea
\delta B&=&\sqrt{5.3\,B_{Z+j}+ \sum_i B_{i}+\sum_i (0.1B_{i})^2},\label{eq:error}\\
&&i=t\bar t\rm{, } \,W(\rightarrow \ell\nu_\ell)+j \rm{, }\,W(\rightarrow \tau \nu_\tau)+j \nonumber,
\eea
where we assign an overall systematic uncertainty of $10\%$ for all backgrounds apart for 
$Z(\rightarrow \nu \bar\nu)+j.$\footnote{One can measure $Z(\rightarrow e^+e^-/\mu^+\mu^-)+j$. From the known $Z$ branching ratios one can estimate the background cross section. However, the statistical error will be increased and including the efficiency of the calibration sample $Z(\rightarrow e^+e^-/\mu^+\mu^-)+j$ in the signal region, the sample is roughly a factor of 5.3 smaller than $Z(\rightarrow \nu \bar\nu)+j$ \cite{Vacavant}.} We see that the discovery of stop pairs in association with a jet should be possible for stop masses up to 290 GeV and for mass splittings between the stop and neutralino of up to 45 GeV. Stop masses up to 360 GeV can be excluded at $2\sigma$ if the mass splitting is very small. With increasing mass splitting, the significance gets worse, since the energy distribution of the charm jets becomes harder. This reduces the missing transverse energy and at the same time increases the probability that the signal fails the second jet veto. In the case of large mass splitting, stop pair production can be probed in di--jet plus missing transverse energy signatures. 
In both cases, charm tagging would be useful to suppress the SM background further.
\begin{figure}
\begin{center}
\includegraphics[scale=0.7]{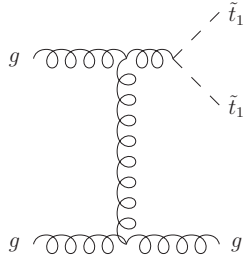}        
\caption{Example diagram for stop pair production in association with one jet via gluon fusion.}
\label{fig:monojet}
\end{center}
\end{figure}

\section{2 $b$--Jets And Missing Transverse Momentum}
\begin{figure}[ht]
\begin{minipage}[b]{0.4\linewidth}
\centering
\includegraphics[scale=0.7]{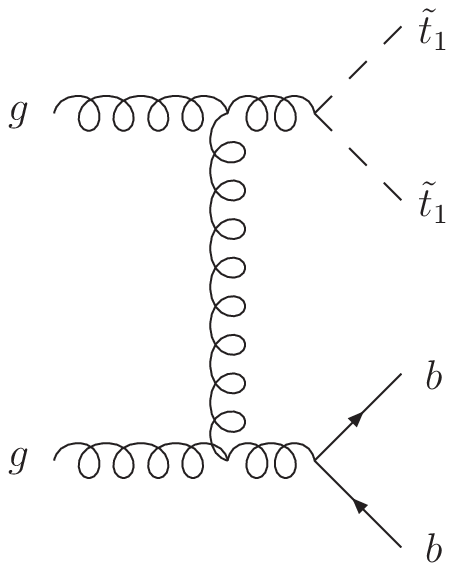}
\caption{Example diagram for QCD stop pair production in association with two
  $b-$jets via gluon fusion.}
\label{fig:qcd_t1t1bb}
\end{minipage}
\hspace{0.5cm}
\begin{minipage}[b]{0.4\linewidth}
\centering
\includegraphics[scale=0.7]{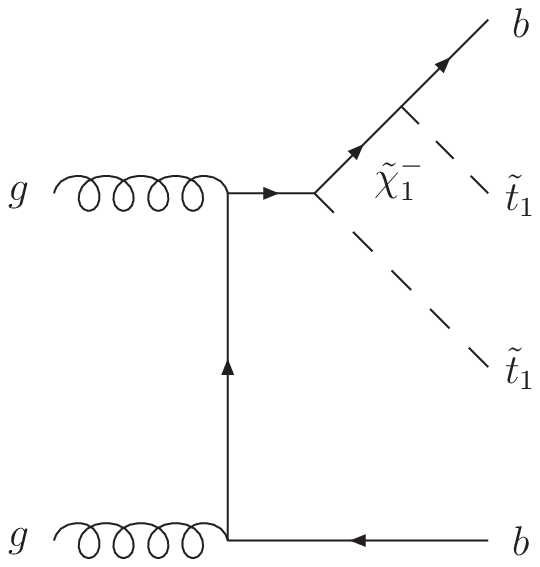}
\caption{Example diagram for mixed QCD--EW stop pair production in association with two
  $b-$jets via gluon fusion. The chargino, $\tilde{\chi}_1^-$, might be
  on--shell.}
\label{fig:ew_t1t1bb}
\end{minipage}
\end{figure}

In the previous section, we discussed the discovery potential of stop pair production with one additional parton. Here, we want to consider an alternative process: stop pair production in association with two $b$--jets \cite{Bornhauser:2010mw} 
\beq
pp\rightarrow \st\st^* b \bar b.
\eeq
It is closely related to the monojet signature as can be seen in Fig. \ref{fig:qcd_t1t1bb}. The charm jets are again too soft to be useful and the hadron collider signature is large missing transverse energy and two $b$--flavoured jets.  We also include the mixed QCD--EW contributions. An example diagram is given in Fig. \ref{fig:ew_t1t1bb}. We assume a neutralino LSP, stop NLSP and a higgsino--like chargino NNLSP. Thus, the resonant decay of the chargino into a stop and a $b$ quark is kinematically allowed and the process depicted in Fig. \ref{fig:ew_t1t1bb} will be a $2\rightarrow3$ process and can give a significant contribution to the pure QCD prediction. However, the size of the mixed QCD--EW contributions depend on the chargino mass and its composition.

The dominant background process is again $Z(\rightarrow \nu \nu)+b \bar b$. The two $b$--jets originate from a $b\bar b$ pair generated via gluon splitting. We also consider the $W(\rightarrow e\nu,\mu\nu,\tau \nu) + b \bar b$ and $t\bar t$ background processes. In addition, we include single top production for our background estimation. 

We require both hard jets ($p_T(b_1)>150$ GeV and $p_T(b_2)>50$ GeV) to be tagged as $b$--jets, since this greatly suppresses the SM backgrounds. In order to sufficiently suppress the QCD background, we demand large missing transverse momentum with $\mpt>200$ GeV. The signal process possesses nearly no isolated leptons and thus we also employ a lepton as well as a tau veto. In addition, we employed cuts on charge multiplicity and the ratio of the $p_T$ of the most energetic $b$--jet and \mpt \cite{Bornhauser:2010mw}. 

\begin{figure}[ht]
\begin{minipage}[b]{0.45\linewidth}
\centering
\includegraphics[width=1.0\textwidth]{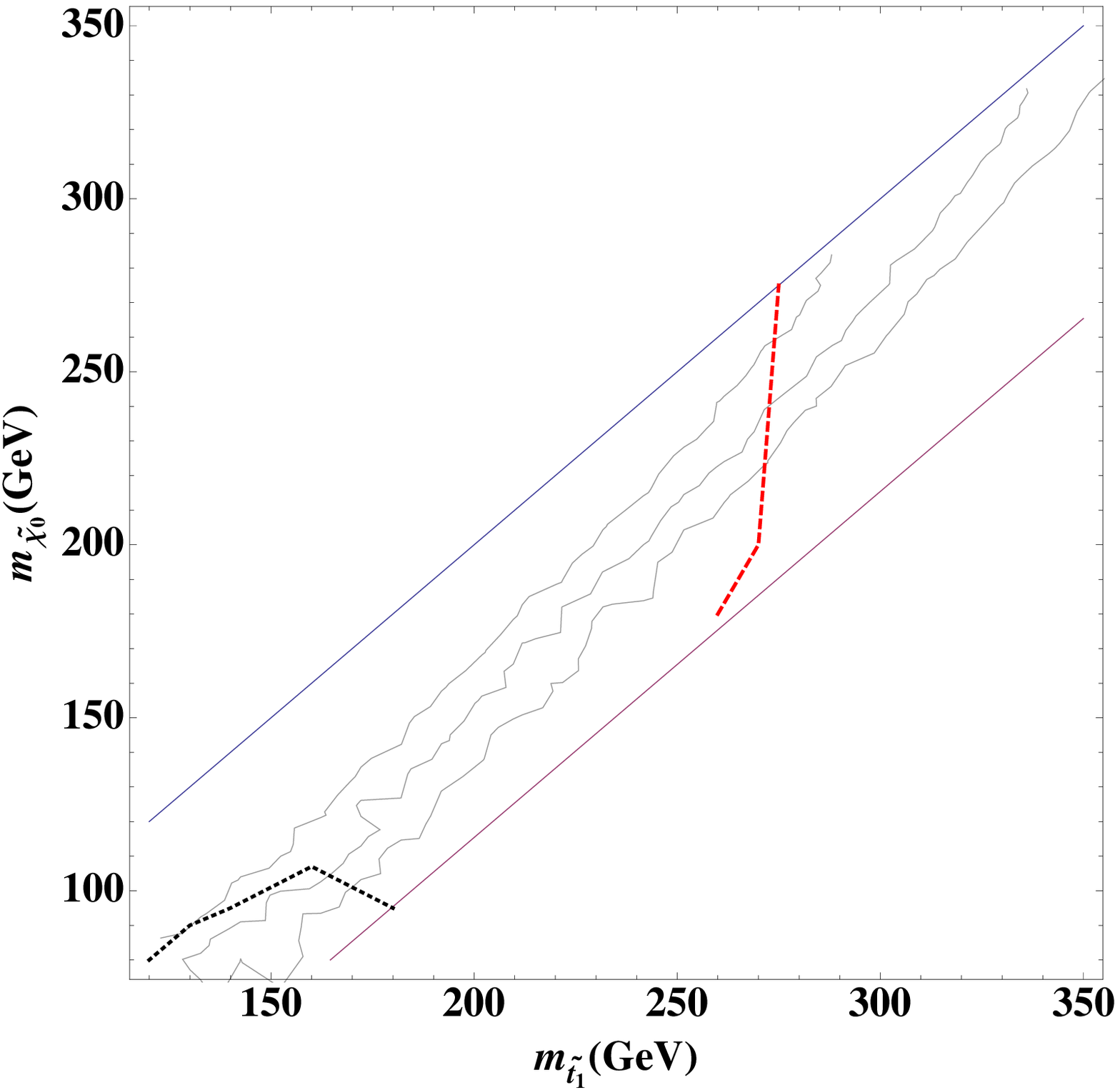}
\caption{Monojet signal significance with background error estimated as in
  Eq.~(2.4)
in the stop--neutralino mass plane assuming an
  integrated luminosity of $100\, \rm{fb}^{-1}$ at $\sqrt{s}=14$
  TeV. The two parallel straight lines delineate the region where
 $\tilde t_1 \rightarrow \tilde \chi_1^0 c$ decays are allowed but
  $\tilde t_1 \rightarrow \tilde \chi_1^0 W^+ \bar b$ decays are
  forbidden. The three grey lines correspond to 5$\sigma$, 3$\sigma$
  and 2$\sigma$ (from top to bottom), respectively. The short-dashed
  black curve delimits the Tevatron exclusion region, whereas the
  long-dashed red curve denotes the upper limit of the discovery reach
  of searches for light stops in events with two $b-$jets and large
  missing energy.}
\label{fig:mstop_mneut_distribution}
\end{minipage}
\hspace{0.5cm}
\begin{minipage}[b]{0.45\linewidth}
\centering
\includegraphics[width=1.0\textwidth]{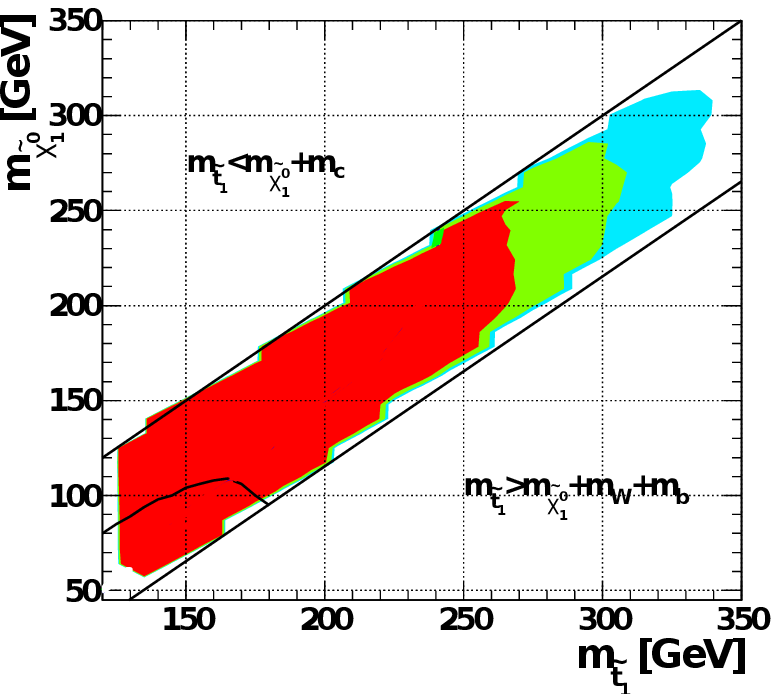}
\caption{Statistical significance of the $\tilde t_1\tilde t_1^* b\bar b$ at the LHC with $\sqrt{s}=14$ TeV as
  a function of the stop and neutralino mass. The red, green and turquoise region
   corresponds to an excess of at least 5$\sigma$, 3$\sigma$ and 2$\sigma\
$, respectively,
  for an integrated luminosity of $100\,\rm{fb}^{-1}$. The chargino mass is fixed by
  $m_{\tilde\chi_1^\pm}=m_{\tilde t_1}+20$ GeV. The parameter space {\em below} the black
  curve is excluded by Tevatron searches \cite{CDFnote9834}. The parameter region where $\tilde t_1$ decays
  into a charm and a neutralino are expected to dominate is given by the
  condition $m_{\tilde{\chi}_1^0}+m_c < m_{\tilde{t}_1} < m_{\tilde{\chi}_1^0}  
  + m_W + m_b$. }
\label{fig:mstop_mneut_mchargconst}
\end{minipage}
\end{figure}

In Fig. \ref{fig:mstop_mneut_mchargconst}, we show the significance in the stop--neutralino mass plane for an integrated luminosity of 100 fb$^{-1}$ at $\sqrt{s}=14$ TeV. Contrary to the monojet signal study, we did not include systematic errors and the statistical significance is only given by $S/\sqrt{B}$, where $S$ and $B$ denotes the number of signal and backgrounds events, respectively. The region below the black curve is ruled out by the Tevatron searches at $95\%$ confidence level. The significance increases with decreasing stop--neutralino mass difference. This is similar to the monojet signal and the arguments are analogous. Our numerical results show that the discovery of stops with masses up to $m_{\st}\le270$ GeV is possible. Stops masses up to $m_{\st}\le340$ GeV can be excluded. 

This analysis is not comparable to the results from the monojet search channel. First, we estimated the significance by including the statistical uncertainties only. Secondly, detector effects were not included in the 2 $b$--jets and missing transverse momentum channel. In addition, the monojet signature does not depend on the higgsino--like chargino mass and hence the significance presented in Fig. \ref{fig:mstop_mneut_mchargconst} is more model dependent.

\section{Coupling reconstruction}
As we have seen in the previous section, it might be possible to measure the rate of the process $pp\rightarrow \st\st^*b\bar b$. The final state is interesting, because the mixed QCD--EW contributions in Fig. \ref{fig:ew_t1t1bb} are sensitive to the coupling $\lambda$. Therefore, we can find an estimate of this coupling, which provides a test of the respective SUSY coupling relation and thus a test of supersymmetry itself. In this talk, we only present a {\it rough} estimate of the precision with which the coupling $\lambda$ can be reconstructed assuming an integrated luminosity of 100 fb$^{-1}$ at $\sqrt{s}=14$ TeV.\footnote{A much more detailed study will be presented in the near future.}. The measured rates depend on the stop, neutralino and the chargino mass and $\lambda$. However, the monojet signal only depends on the stop and neutralino mass. By measuring the monojet event rates and the missing transverse momentum distribution, we can determine the stop and neutralino mass with a high precision.  The stop mass itself uniquely determines the $\st\st^*b\bar b$ QCD contributions. Thus, we can estimate the magnitude of the mixed QCD--EW diagrams if we know the chargino mass and $\lambda$. Assuming that we can determine the pure QCD $\st\st^*b\bar b$ contribution, we can calculate the mixed QCD--EW contribution. The latter LO cross section is proportional to the square of the $\lambda$ coupling, thus the relative error of the coupling is roughly 1/2 of the relative error of the mixed QCD--EW cross section. The chargino mass can be obtained by the cross section and shape comparisons of kinematical distributions such as the $p_T$ distribution of the $b$--jet or the difference in rapidity between the leading two $b$--jets.  Here, we consider a scenario with $m_{\tilde\chi_1^0}=180\,\rm{GeV}$, $m_{\tilde t_1}=200\,\rm{GeV}$ and $m_{\tilde\chi_1^\pm}=220\,\rm{GeV}$. In the following, we briefly discuss the main systematic and statistical errors.

In order to translate the number of reconstructed events in a cross section, the total integrated luminosity is needed. We assume that the luminosity can be measured with a precision of $3\%$.

The absolute cross section depends on the parton distribution function (PDF). Since the largest contribution to the process comes from gluon initial states, the uncertainty of the gluon PDF gives the largest contribution to the error. We estimated the error by comparing the events rates with different PDF sets (CTEQ6l, CTEQ5l, MSTW2008nlo) and obtain an error of about 8\%.

Unfortunately, our process is only known at leading order (LO) and thus the next--to--LO (NLO) corrections might be sizeable. The uncertainty of those missing corrections are estimated by varying the factorisation and renormalisation scale from $1/2\,m_{\st}$ to $2\,m_{\st}$ leading to an error of about 50\%. However, this is a very conservative estimate, since NLO calculations would be certainly known, if such a measurement were performed.

We estimate the precision of the mass reconstruction with $\Delta m=5$ GeV. This leads to a relative error on the cross section of about 10\%.
The SM background's statistical fluctuations contribute to the error. We estimated the error with $\sqrt{B}$ and obtain an error of the order of 17\%.

If we assume that all errors are uncorrelated for the sake of simplicity, we can add all errors in quadrature. We find that the mixed QCD--EW contribution can be determined with an relative error of $50\%$. This results in a relative error of $28\%$ of the coupling $\lambda$. Clearly, the total error is dominated by the uncertainty of the missing NLO corrections. However, higher order calculations would certainly be performed in the future, if an excess above the SM expectation is measured, which is compatible with our collider signature. This would allow us for a much more precise determination of the coupling.
\begin{table}[t!]
\begin{center}
\begin{tabular}{c||c|c}
error & $\Delta \sigma_{\rm{mixed\,QCD-EW}}/\sigma_{\rm{mixed\,QCD-EW}}$ & $\Delta\lambda/\lambda$ \\
\hline
luminosity & 3$\%$ & 1.5$\%$ \\
PDF uncertainty & 8$\%$ & 4$\%$\\
NLO corrections & 50$\%$ & 25$\%$ \\
$\Delta \tilde m=5$ GeV & 14$\%$ & 7$\%$ \\
statistics & $17\%$ & $8.5\%$\\\hline
$\Sigma$& &$28\%$\\
\end{tabular}
\caption{Relative errors for the $\tilde t_1\tilde t_1^*b\bar b$ mixed QCD--EW cross section and the $\st-\charg-b$ coupling.} 
\end{center}
\label{tab:errors}
\end{table}
\section{Conclusion}
In this talk, we presented two search channels for light stops with a small difference between the lighter stop and the neutralino LSP. In such scenarios, simple stop pair production does not yield an observable signal. We first discussed the prospect of discovery of such stops for the monojet signature   before analysing the stop pair production in association with two $b$--quarks.
We presented the discovery reach of both search channels in the stop--neutralino mass plane with an integrated luminosity of 100 fb$^{-1}$ at $\sqrt{s}=14$ TeV. The latter process allows us to probe the $\st-\charg-b$ coupling, thereby allowing to check a SUSY coupling relation involving Yukawa couplings. We presented a rough estimate for measuring this coupling at the LHC.


\begin{thebibliography}{99}
\bibitem{ATLAS_search}
ATLAS-CONF-2012-109, http://cdsweb.cern.ch/record/1472710

\bibitem{Aad:2012yr}
  G.~Aad {\it et al.}  [ATLAS Collaboration],
  arXiv:1209.2102 [hep-ex].

\bibitem{Carena:2008mj} 
  M.~Carena, A.~Freitas and C.~E.~M.~Wagner, 
  JHEP {\bf 0810 } (2008)  109 

\bibitem{Drees:2012dd}
  M.~Drees, M.~Hanussek and J.~S.~Kim,
  Phys.\ Rev.\ D {\bf 86} (2012) 035024
  
\bibitem{Hikasa:1987db} 
  K.~i.~Hikasa and M.~Kobayashi, 
  Phys.\ Rev.\  D {\bf 36} (1987) 724;

\bibitem{Muhlleitner:2011ww}
M.~M\"uhlleitner and E.~Popenda,
JHEP {\bf 1104} (2011) 095 

\bibitem{atlasConf} 
ATLAS Collab., G.~Aad {\it et al.}, 
ATLAS-CONF-2010-065. 

\bibitem{Vacavant} 
  L.~Vacavant and I.~Hinchliffe, 
  J.\ Phys.\ G {\bf G27 } (2001)  1839. 
  
\bibitem{Bornhauser:2010mw}
  S.~Bornhauser, M.~Drees, S.~Grab and J.~S.~Kim,
  Phys.\ Rev.\ D {\bf 83} (2011) 035008
  
 \bibitem{CDFnote9834}
  [CDF Collaboration],
  CDF note 9834.

\end{thebibliography}
\end{document}